\documentclass{PoS}

\usepackage[utf8x]{inputenc}
\usepackage{amsmath}
\usepackage{nicefrac}
\usepackage[merge,sort&compress,numbers]{natbib}
\usepackage{url}
\usepackage{subcaption}
\usepackage{soul}

\title{Anatomy of SU(3) flux tubes at finite temperature}

\author{Paolo Cea\\
        Dipartimento di Fisica, Universit\`a di Bari, \\
        and INFN - Sezione di Bari, I-70126 Bari, Italy\\
        E-mail: \email{paolo.cea@ba.infn.it}}
\author{Leonardo Cosmai\\
        INFN - Sezione di Bari, I-70126 Bari, Italy\\
        E-mail: \email{leonardo.cosmai@ba.infn.it}}
\author{\speaker{Francesca Cuteri}\\
        Dipartimento di Fisica, Universit\`a della Calabria, \\
        and INFN - Gruppo collegato di Cosenza, I-87036 Rende, 
        Cosenza, Italy\\
        E-mail: \email{francesca.cuteri@fis.unical.it}}
\author{Alessandro Papa\\
        Dipartimento di Fisica, Universit\`a della Calabria, \\
        and INFN - Gruppo collegato di Cosenza, I-87036 Rende, 
        Cosenza, Italy\\
        E-mail: \email{alessandro.papa@fis.unical.it}}

\ShortTitle{Anatomy of SU(3) flux tubes at finite temperature}

\abstract{An attempt to adapt the study of color flux tubes to the case of 
finite temperature has been made.
The field is measured both through the correlator of two Polyakov loops, one of 
which connected to a plaquette, and through a connected correlator of Wilson 
loop and plaquette in the spatial sublattice. Still the profile of the 
flux tube resembles the transverse field distribution around an isolated vortex 
in an ordinary superconductor.
The temperature dependence of all the parameters characterizing the flux tube is
investigated.}

\FullConference{The 33rd International Symposium on Lattice Field Theory\\
		14 -18 July 2015\\
		Kobe International Conference Center, Kobe, Japan}

\begin{document}

\section{Introduction}

The color confinement phenomenon accounts for our inability to detect free 
colored particles, with the QCD spectrum consisting of color-singlet
particle states only. Although yet unexplained from first principles, color 
confinement can be interpreted within a few possible scenarios and lattice QCD 
studies are relevant in order to verify/disprove the validity of 
the different options.

It was conceived, by 't Hooft and Mandelstam, that the QCD vacuum could 
be modeled as a coherent state of color magnetic monopoles, called 
\emph{dual superconductor}, since the condensation of color magnetic monopoles 
is thought to be analogous to the formation of Cooper pairs in the BCS theory of
superconductivity~\cite{'tHooft:1976ep,*Mandelstam:1974pi,*Ripka:2003vv}. A dual
superconductor is a superconductor in which the roles of the electric and magnetic
fields are exchanged.
The analogy is suggested both by the the absence of free colored states, and by 
the fact that meson resonances lie approximately on Regge trajectories, 
indicating that a quark-antiquark $q\bar{q}$ pair is connected by a string
with a constant string tension, i.e. with an energy that increases 
linearly with the distance $R$ between the color charges. Correspondingly, in 
the large-distance regime, the nonperturbative dynamics squeezes the 
color-field lines, giving rise to flux tubes connecting the two charges. 
The formation of color flux tubes can be interpreted as the dual analog of the 
Meissner effect.
Lattice QCD allows us to investigate the color confinement phenomenon 
nonperturbatively and, in this framework, convincing evidences both for 
the color magnetic monopole condensation, and for the existence of color 
tubelike structures, have been produced~\cite{Bander:1980mu,*Fukugita:1983du,*Kiskis:1984ru,*Flower:1985gs,*Wosiek:1987kx,*DiGiacomo:1990hc,*Cea:1992vx,*Singh:1993jj,*Cea:1993pi,*Cea:1994ed,*Shiba:1994db,*Bali:1994de,*Arasaki:1996sm,*DiGiacomo:1999fb,*Cea:2000zr,*Cea:2001an,*Carmona:2001ja,*Greensite:2003bk,*Cea:2004ux,*Haymaker:2005py,*D'Alessandro:2006ug,*D'Alessandro:2010xg}.
In previous studies~\cite{Cea:1995zt,*Cardaci:2010tb,*Cea:2012qw,*Cea:2014uja} 
the tubelike distribution of color fields in presence of static quarks has been 
studied both in the SU(2) and SU(3) pure gauge theory at zero temperature.
As a meaningful extension of those studies, the structure of flux tubes in the
SU(3) pure gauge theory at nonzero temperature, and across the deconfinement transition
temperature $T_c$, is now investigated.
There is a twofold motivation for this study: on one hand, the nonperturbative 
study of flux tubes at $T\neq0$ is relevant to clarify the formation of $c\bar{c}$ and
$b\bar{b}$ bound states in heavy-ion collisions at high energies; on the other hand,
the study of the behavior of flux tubes across $T_c$ allows us to check the validity of
the dual superconductor model.
For our investigation we have made use of the MILC code, which has been suitably
modified by us in order to introduce the relevant observables. The use of the 
MILC code will allow, in future, simulations for the physically relevant 
case of full QCD with dynamical quarks.

The color field distribution generated by a $q\bar{q}$ static pair in the 
vacuum is probed by means of the connected correlators~\cite{DiGiacomo:1990hc,*DiGiacomo:1989yp,*Kuzmenko:2000bq,*DiGiacomo:2000va}
\begin{figure}[!htb]
  \begin{subfigure}[b]{0.55\textwidth} \centering
    \includegraphics[width=0.8\columnwidth,height=5cm,clip]{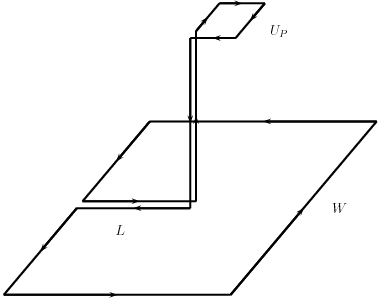}
  \end{subfigure}
  \hfill
  \begin{subfigure}[b]{0.35\textwidth} \centering 
    \includegraphics[width=\columnwidth,height=5cm,clip]{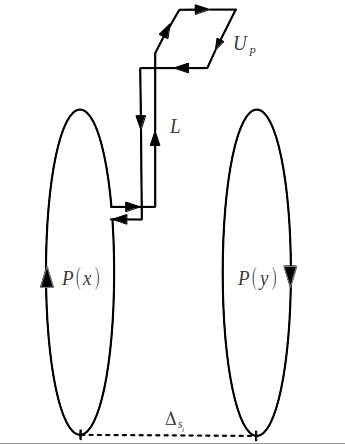}
  \end{subfigure}
  \caption{\label{fig:correlators}(Left) The connected correlator between the 
  plaquette $U_P$ and the Wilson loop. (Right) The correlator between two 
  Polyakov loops, one of which connected to a plaquette $U_P$. The subtraction 
  in $\rho_{P,W}^{\rm conn}$ is not explicitly drawn.}
\end{figure}
\begin{equation}
\label{eq:rhoPconn}
\rho_{P}^{\rm conn} = \frac{\left\langle \mathrm{tr}\left(P\left(x\right)LU_{P}
L^{\dagger}\right)\mathrm{tr}P\left(y\right)\right\rangle }{\left\langle 
\mathrm{tr}\left(P\left(x\right)\right)\mathrm{tr}\left(P\left(y\right)\right)
\right\rangle } - \frac{1}{3}\frac{\left\langle \mathrm{tr}\left(P\left(x\right)
\right)
\mathrm{tr}\left(P\left(y\right)\right)\mathrm{tr}\left(U_{P}\right)\right
\rangle }{\left\langle \mathrm{tr}\left(P\left(x\right)\right)\mathrm{tr}
\left(P\left(y\right)\right)\right\rangle}\;
\end{equation}
and
\begin{equation}
\label{eq:rhoWconn}
\rho_W^{\rm conn} = \frac{\left\langle {\rm tr}
\left( W L U_P L^{\dagger} \right)  \right\rangle}
              { \left\langle {\rm tr} (W) \right\rangle }
 - \frac{1}{N} \,
\frac{\left\langle {\rm tr} (U_P) {\rm tr} (W)  \right\rangle}
              { \left\langle {\rm tr} (W) \right\rangle } \;.
\end{equation}
In both cases, $U_P=U_{\mu\nu}(x)$ is the plaquette in the $(\mu,\nu)$ plane, and 
$L$ is the Schwinger line connecting the plaquette, either to the Wilson loop 
$W$, or to a Polyakov loop $P$; $N$ is the number of colors. A schematic 
representation of the correlators is given in Fig.~\ref{fig:correlators}.
The use of the connected correlator~\eqref{eq:rhoWconn} at $T\neq0$ is limited 
to the exploration of the chromomagnetic sector.
The linearity of $\rho_{W}^{conn}$ and $\rho_{P}^{conn}$ in the field, in the SU(3) 
case, holds up to terms of order $a^2$ in the lattice spacing.
The linear term in $F_{\mu\nu}$, then, dictates the dominant behavior of our 
correlators in the continuum limit, and a symbolic expression for the naive 
continuum limit of the connected correlators is 
\begin{equation}
\label{rhoWlimcont}
\rho_{W,P}^{\rm conn}\stackrel{a \rightarrow 0}{\longrightarrow} a^2 g 
\left[ \left\langle
n^aF^a_{\mu\nu}\right\rangle_{q\bar{q}} \right],
\end{equation}
In this formula $\langle\quad\rangle_{q \bar q}$ denotes the ensemble 
average of the field projection onto an unknown direction (in color space), 
$n^a$, determined by the static $q \bar q$ pair, and 
$\beta = \nicefrac{2N}{g^2}$ is the coupling constant.
The components of the field strength tensor can, then, be extracted as
\begin{equation}
F^a_{\mu\nu}\left(x\right)n^a=\sqrt{\frac{\beta}{2N}}\rho_{W,P}^{conn}\left(x\right)
\;.
\end{equation}

The color field distribution of flux tubes and, in particular, the shape of 
color fields, in a direction perpendicular to the axis of the flux tube joining 
the $q \bar q$ pair, is probed by varying position and orientation of the 
plaquette.
The dual superconductor model enters our analysis in the choice of the function 
to fit the transverse shape of the field. Here we exploit the dual analog of 
a result presented in~\cite{Clem:1975aa}:
%
\begin{equation}
E_{l}\left(x_{t}\right)=\frac{\phi}{2\pi}\frac{\mu^{2}}{\alpha}\frac{K_{0}\left[\left(\mu^{2}x_{t}^{2}+\alpha^{2}\right)^{\nicefrac{1}{2}}\right]}{K_{1}\left[\alpha\right]} =  \frac{\phi}{2 \pi} \frac{\mu^2}{\alpha} \frac{K_0[(\mu^2 x_t^2 
+ \alpha^2)^{1/2}]}{K_1[\alpha]}\;,\label{eq:clem}
\end{equation}
where $\mu=\nicefrac{1}{\lambda}$ is the inverse of the London penetration 
depth, $\xi_v$ is the variational core radius parameter, and the quantity at 
the left-hand side is the longitudinal chromoelectric field, which has 
been found (also at $T\neq0$) to be the only statistically sizable color field 
component forming the flux tube.
By fitting Eq.~\eqref{eq:clem} to $E_l(x_t)$ data, one can extract $\phi$, 
$\lambda$ and $\xi_v$. The Ginzburg-Landau $\kappa$ parameter can be obtained by
\begin{equation}
\label{landaukappa}
\kappa = \frac{\lambda}{\xi} =  \frac{\sqrt{2}}{\alpha} 
\left[ 1 - K_0^2(\alpha) / K_1^2(\alpha) \right]^{1/2} \;,
\end{equation}
and the coherence length $\xi$ can be deduced from $\kappa = 
\nicefrac{\lambda}{\xi}$.

\section{Color flux tubes on the lattice}
We performed our numerical simulations using the Wilson action on 
lattices with periodic boundary conditions and the heat-bath algorithm 
combined with overrelaxation. To reduce the autocorrelation time, measurements 
were taken every 10 updatings.
The considered lattice sizes had temporal extensions ranging from $L_t=10$ up 
to $L_t=16$ and spatial size $L_s$ fixed to keep the aspect ratio equal to 
four. The temperature varies according to:
\begin{equation}
\label{temperature}
T \; = \; \frac{1}{a(\beta) \, L_t} \; ,
\end{equation}
where the scale is fixed with the parameterization in~\cite{Edwards:1997xf} and 
$\sqrt{\sigma} = 420\,\text{MeV}$.
A relevant variable to be taken under control in our simulations is the distance
$\Delta$ between the static sources (size of the Wilson loop or distance 
between Polyakov loops).
Consistently with our previous studies~\cite{Cea:2014uja}, we have found that 
$\Delta$ has to be chosen in a way that the distance in physical units is 
beyond half a fermi. In such a distance regime, the only parameter affected by 
changes in the physical distance, produced for example by keeping $\Delta$ 
fixed in lattice units, while varying $\beta$, is $\phi$.
It was then decided to approximately fix the physical, rather than the lattice, 
distance between the sources. The advantages are that less statistics is needed 
for smaller couplings and that the achievement of the continuum limit can 
be shown already at the level of the measured field.
In order to reduce the ultraviolet noise, we applied one step of HYP 
smearing~\cite{Hasenfratz:2001hp} to links in the temporal direction, 
with smearing parameters $(\alpha_1,\alpha_2,\alpha_3) = (1.0, 0.5, 0.5)$, and 
$N_{\rm APE}$ steps of APE smearing~\cite{Albanese:1987ds} to spatial 
links, with smearing parameter $\alpha_{\rm APE} = 0.50$.
As a criterion to determine the optimal smearing step, the position of the peak 
in $\phi$ (signalling the maximal disentanglement of our signal from background 
noise) has been used.
\begin{figure}[tb] 
\begin{subfigure}[b]{0.5\textwidth}\centering
\includegraphics*[width=0.95\columnwidth,height=6.5cm,clip]{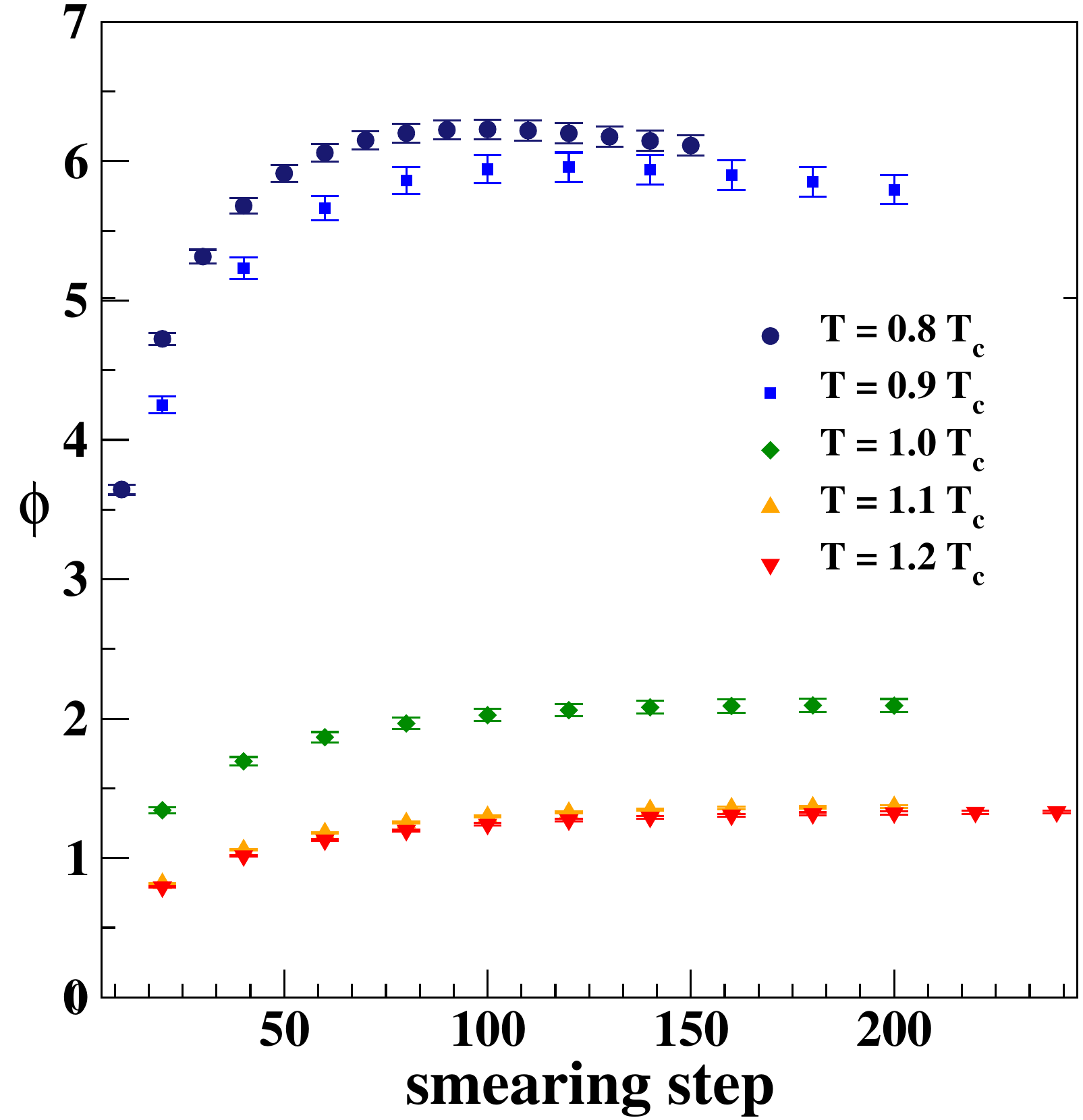}
\label{fig:phi_vs_smear}
\end{subfigure}
\begin{subfigure}[b]{0.5\textwidth}\centering
\includegraphics*[width=0.95\columnwidth,height=6.5cm,clip]{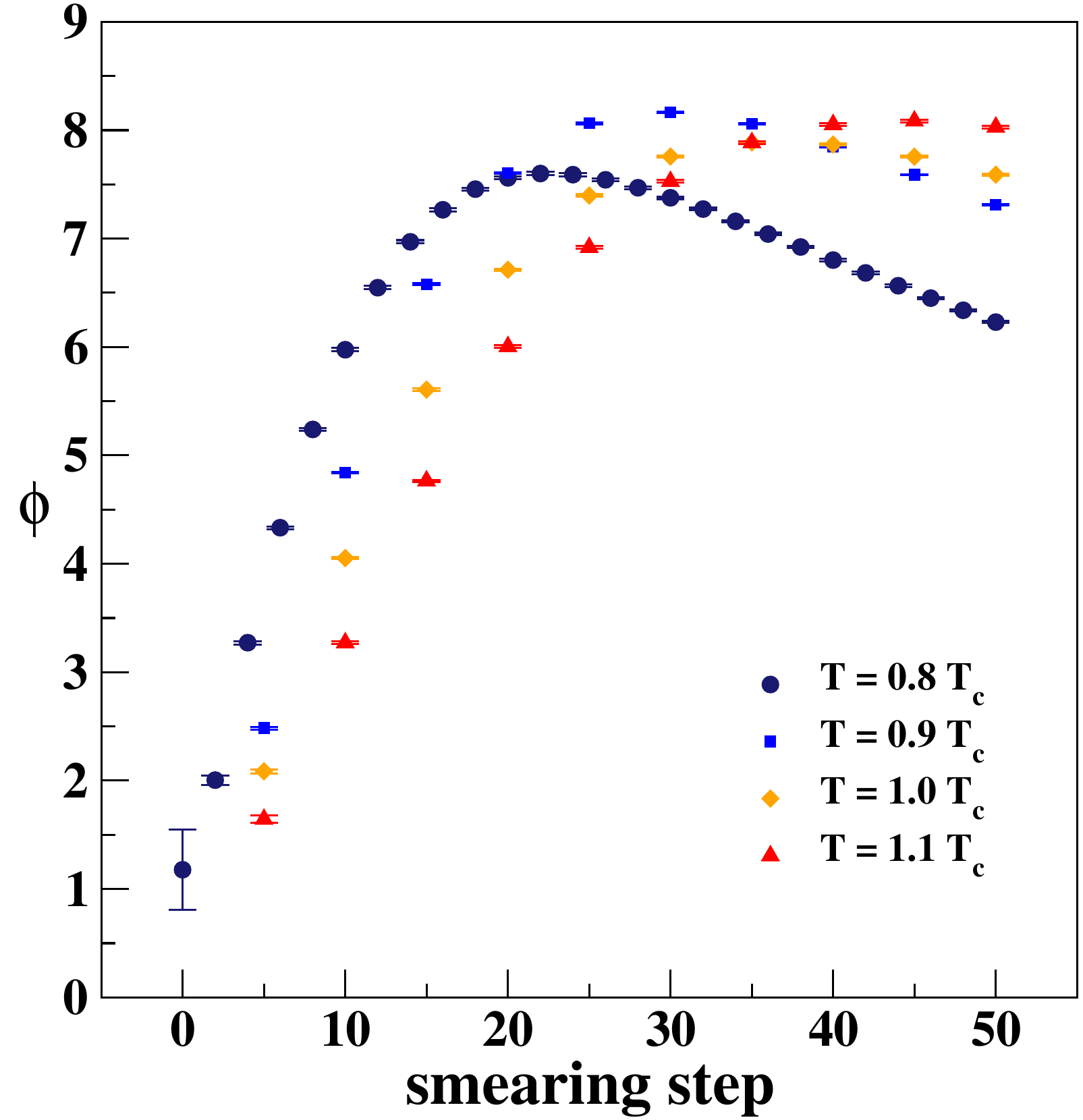}
\label{fig:magnphi_vs_smear}
\end{subfigure}
\caption{$\phi$ {\it vs} smearing for measurements of the Polyakov (left) and Wilson (right) connected correlator at $T\neq0$, on a $40^3\times10$ lattice.}
\end{figure}
%
\section{Flux tubes across the deconfinement transition}

\begin{figure}[tb]
\begin{subfigure}[b]{0.5\textwidth}
\includegraphics[width=\columnwidth,height=6.cm]{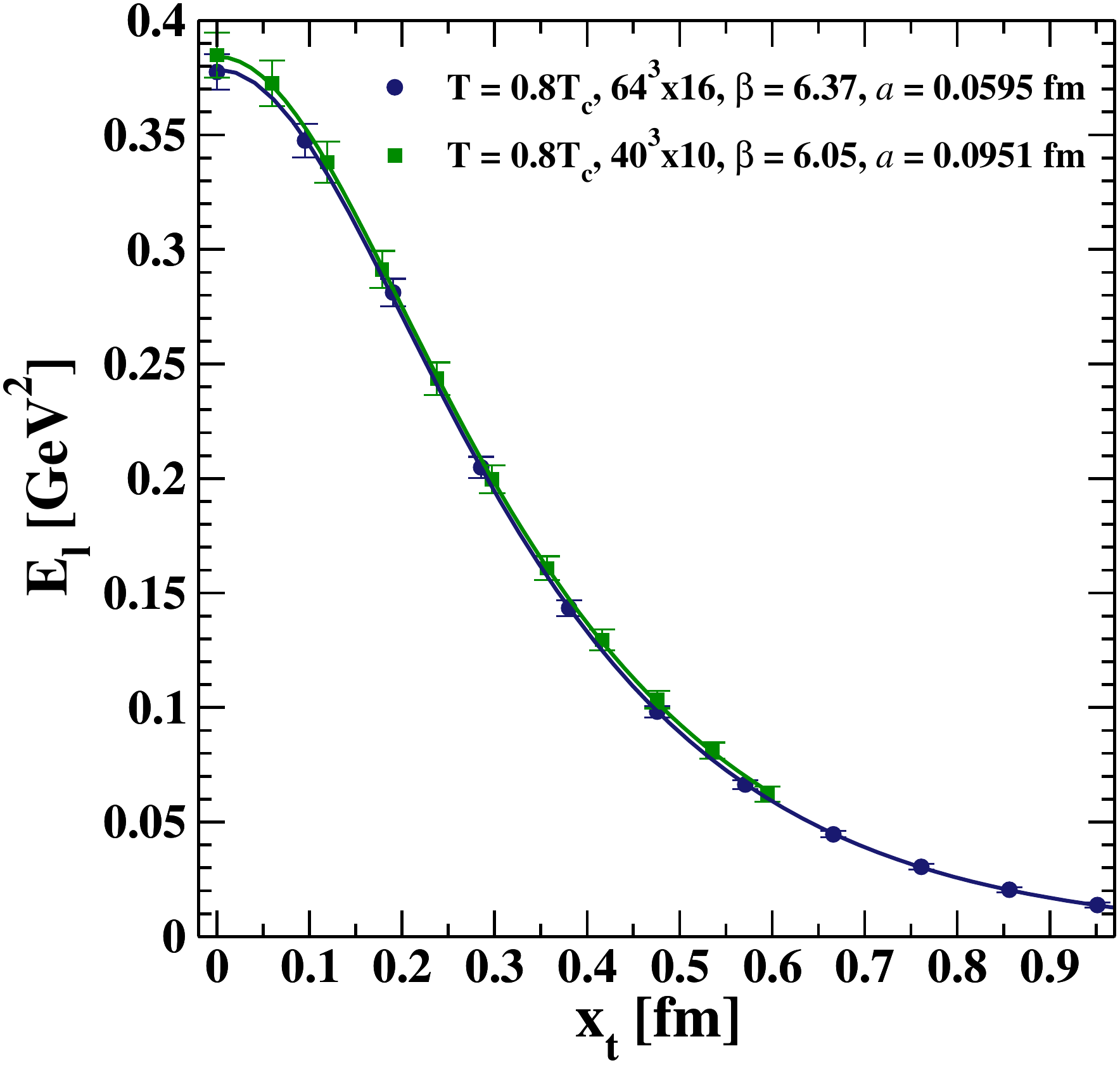}
\end{subfigure}
\begin{subfigure}[b]{0.5\textwidth}
\includegraphics[width=\columnwidth,height=6.cm]{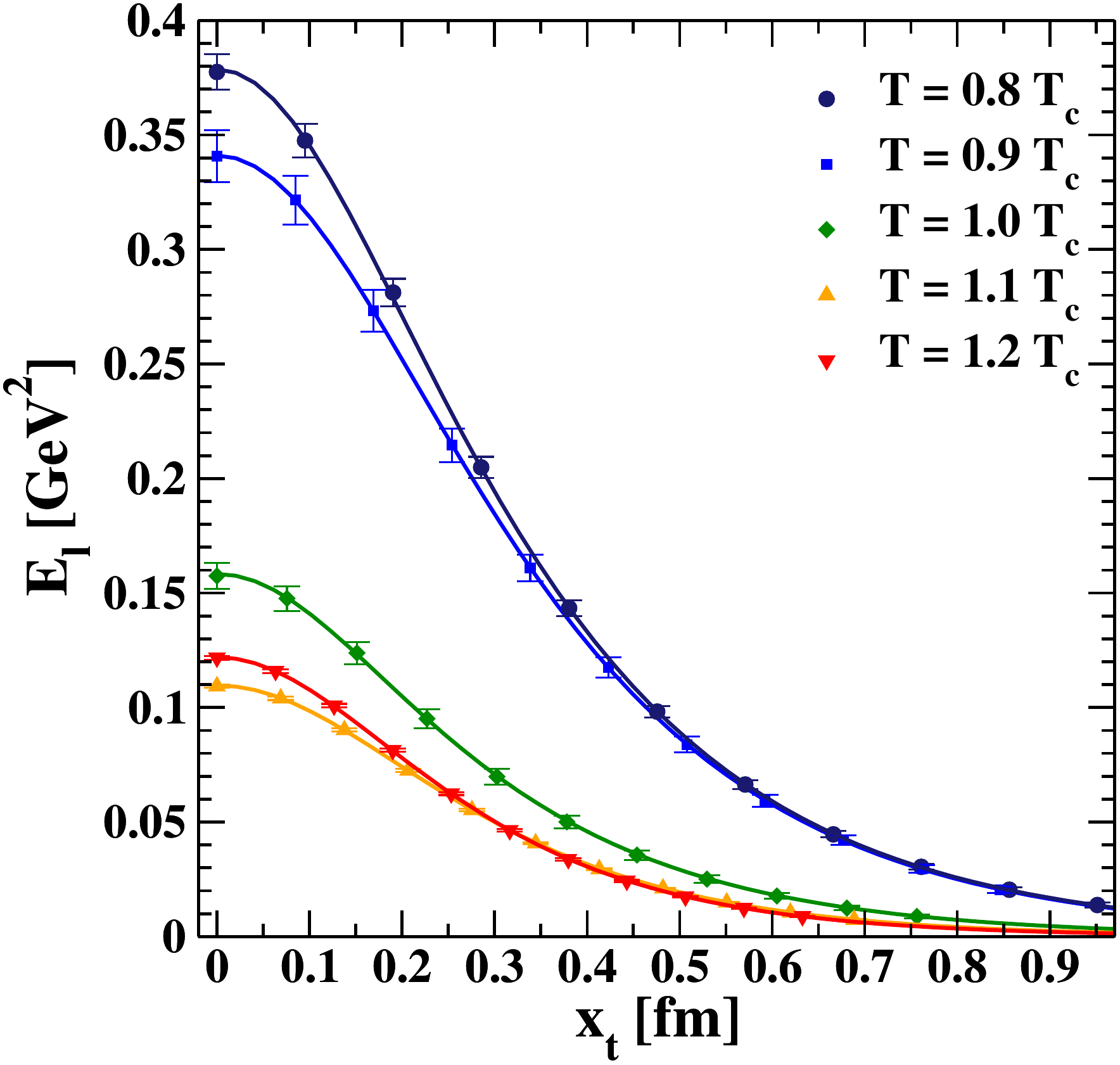}
\end{subfigure} 
\caption{\label{fig:ElField} (Left) Scaling of $E_l\left(x_t\right)$ at $T\simeq
 0.8 T_c$. (Right) $E_l\left(x_t\right)$ at fixed lattice size $40^3\times 10$, 
and couplings $\beta=6.050, 6.125, 6.200, 6.265, 6.325$ corresponding to 
temperatures $T/T_c=0.8, 0.9, 1.0, 1.1, 1.2$. For the critical temperature, we 
used $T_c=260\,\text{MeV}$. The solid lines are the fit of our data to 
Eq.~\protect\eqref{eq:clem}.}
\end{figure}
\begin{figure}[tb]
\begin{subfigure}[b]{0.5\textwidth}\centering
\includegraphics*[width=0.95\columnwidth,height=5.5cm,clip]{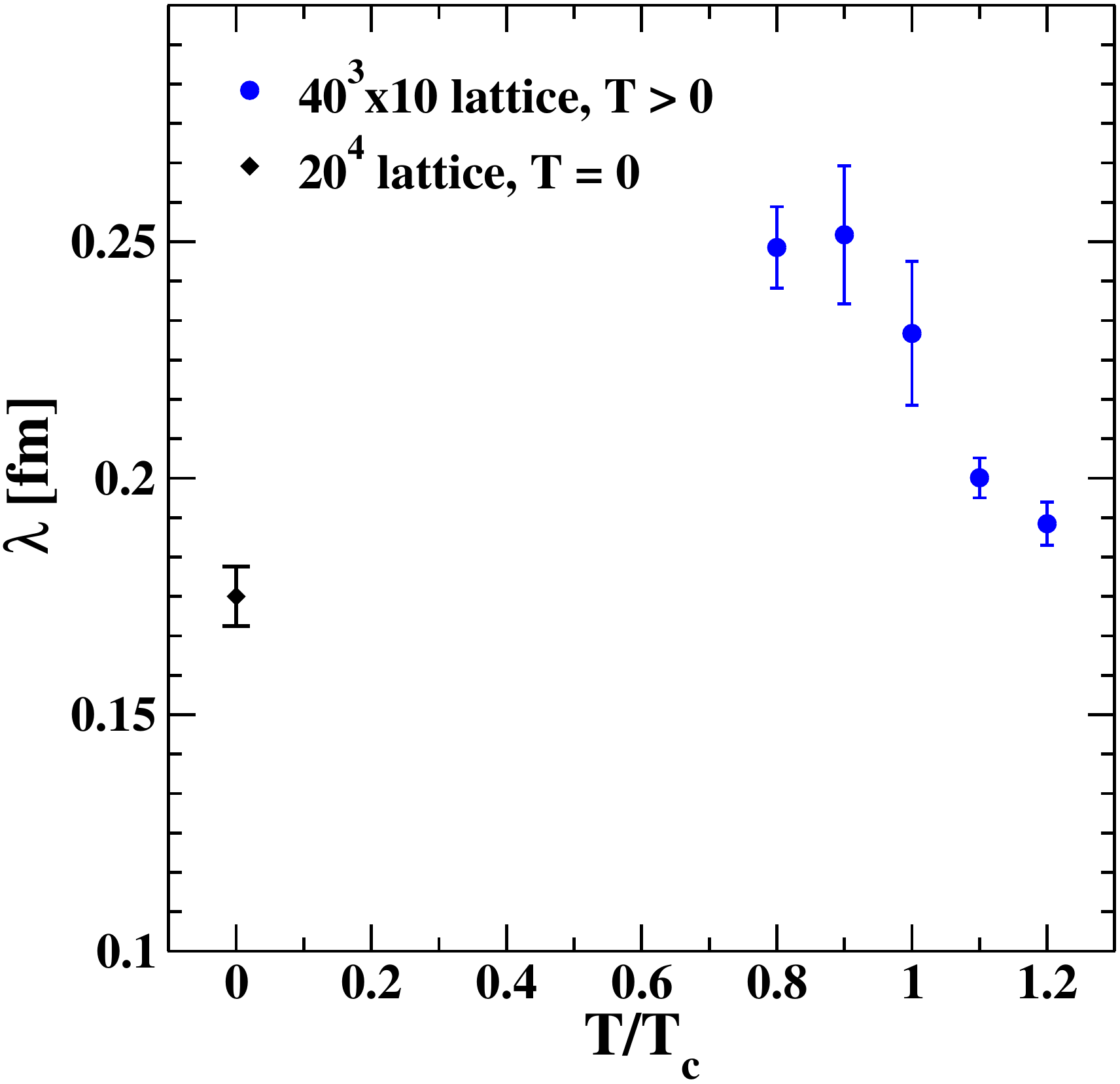}
\end{subfigure}
\begin{subfigure}[b]{0.5\textwidth}\centering
\includegraphics*[width=0.95\columnwidth,height=5.5cm,clip]{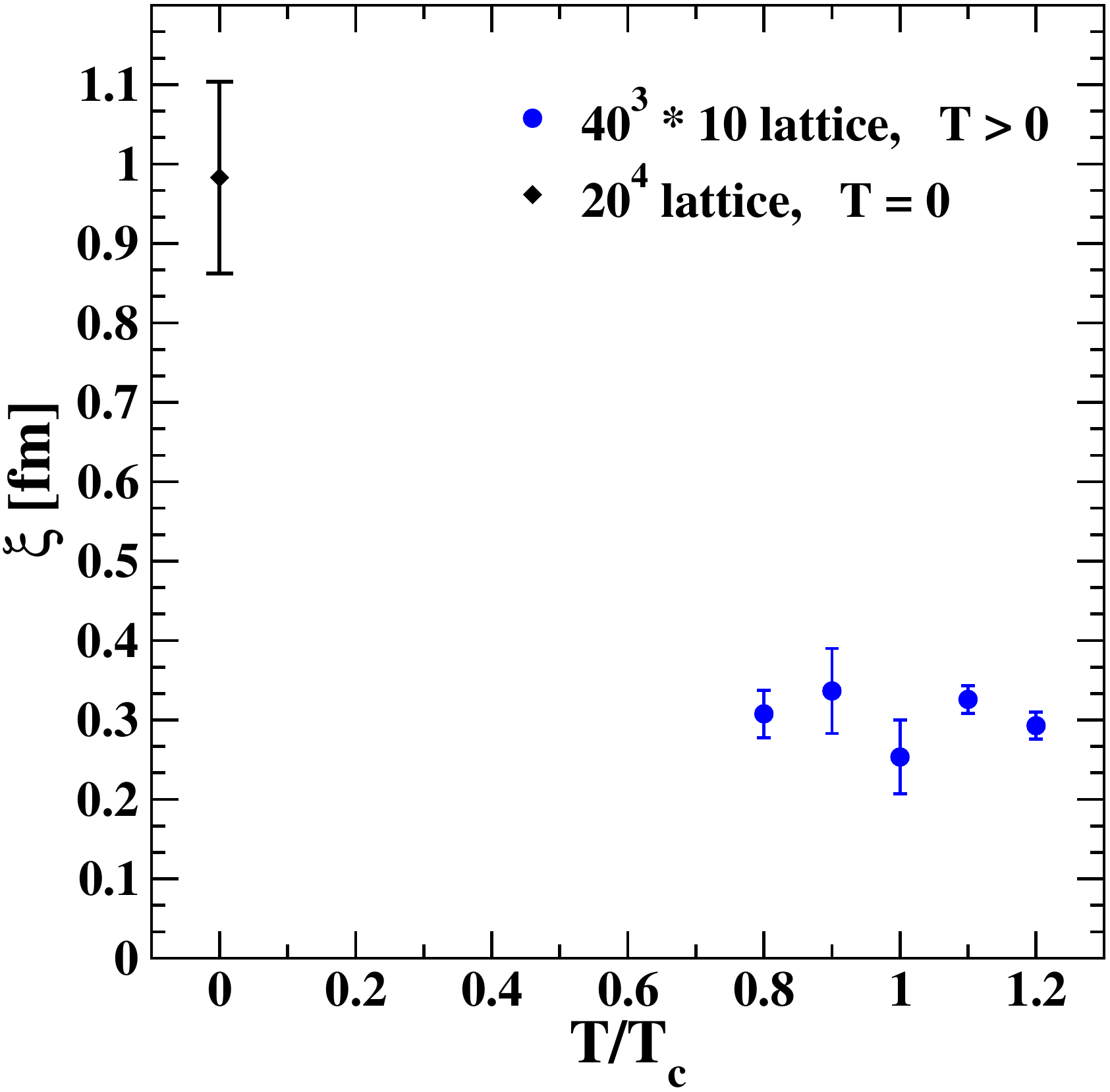}
\end{subfigure}
\caption{(Left) London penetration depth $\lambda$ {\it vs} $T/T_c$ (the 
$\lambda_{T=0}=0.1750(63)$ value is included). (Right) Coherence length 
$\xi$ {\it vs} $T/T_c$ (the $\xi_{T=0}=0.983(121)$ value is included). }
\label{fig:lambdaAndXiVsT}
\end{figure}

In this section we present numerical results on the chromoelectric field 
distribution generated by a static $q\bar{q}$ pair at $T\neq0$, measured 
through the connected correlator in Eq.~\eqref{eq:rhoPconn}.
Before studying the temperature dependence of the chromoelectric field shape, a 
scaling analysis was performed at a fixed temperature of $T=0.8 T_c$.
As shown in Fig.~\ref{fig:ElField} (Left), $E_l\left(x_t\right)$, measured at 
the optimal smearing step, on lattices with different sizes and at $\beta$ 
values tuned in a way to keep the temperature fixed, exhibits the same behavior,
thus indicating continuum scaling.
Afterwards, the lattice size was fixed to $40^3\times 10$ and, by varying the 
coupling $\beta$ in the range $\left[6.050-6.325\right]$, the effect
on the flux tube of a growing temperature was studied. The results 
of our analysis are shown in Fig.~\ref{fig:ElField} (Right).
\begin{table}[tb]
\begin{center} 
\caption{Fit values on a $N_s\times N_t = 40^3\times 10$ lattice for several 
values of $T$.}
\label{tab:fit_summary}
\begin{tabular}{|c|c|c|c|c|c|c|c|}
\hline\hline
$\beta$ & $\Delta$ [fm] & $T/T_c$ & $\phi$ & $\mu$ & $\xi_v$ & $\chi_r^2$ \\ \hline
6.050	&0.761	&0.8	&6.201(68)	&0.382(13)	&3.117(191)	&0.02\\
6.125	&0.761	&0.9	&5.941(101)	&0.337(20)	&3.652(360)	&0.01\\
6.200	&0.756	&1.0	&2.061(45)	&0.328(22)	&3.312(389)	&0.01\\
6.265	&0.757	&1.1	&1.359(9)	&0.344(7)	&4.286(131)	&0.06\\
6.325	&0.760	&1.2	&1.324(11)	&0.332(8)	&4.248(142)	&0.06\\
\hline\hline 
\end{tabular} 
\end{center}
\end{table}
To check to what extent the dual superconductor scenario is confirmed by our 
numerical results, the behavior {\it vs} temperature of
London penetration depth and coherence length was analyzed. Results are shown
in Fig.~\ref{fig:lambdaAndXiVsT}.
We observe that, while the intensity of the measured chromoelectric field has a 
substantial drop at $T_c$, the penetration length $\lambda$ decreases 
monotonically across the transition (but from bigger values than what we 
observed at $T=0$), while $\xi$ stays approximately constant (even though 
at a much smaller value than at $T=0$). According to our numerical findings, 
flux tubes are somehow ``evaporating`` in a way that is not consistent with the 
dual superconductor analogy. In ordinary superconductivity, indeed, both 
$\lambda$ and $\xi$ diverge while $T_c$ is approached.
Nevertheless, flux tubes do survive across the deconfinement transition 
temperature and the field shape is well fitted by the
ansatz derived within the dual superconductivity picture, even beyond $T_c$.

\section{Flux tubes across the deconfinement transition in the magnetic sector}
In the high-temperature regime, through dimensional reduction, QCD can be 
reformulated as an effective three-dimensional theory with the scale of the 
effective couplings given in terms of the temperature~\cite{Kalashnikov:1982sc,*Nieto:1996pi}.  However, straightforward perturbation theory fails in the 
effective theory, even at high-$T$, due to the presence of infrared 
nonperturbative effects which manifest themselves in correlation functions for 
the spatial components of gauge fields.
It is, indeed, known that the spatial Wilson loops obey an area law behavior, 
with spatial string tension $\sigma_s$, also in the high-$T$ phase~\cite{Bali:1993tz,*Karsch:1994af}.
An analysis of the temperature dependence of $\sigma_s$ thus yields information 
on the importance of the nonstatic sector for long-distance properties of 
high-$T$ QCD. For $T\ge2 T_c$ the spatial string tension satisfies:
\begin{equation}
\label{spatial}
\sqrt{\sigma_s}  \; = \; \gamma \; g(T) \; T  \; ,
\end{equation}
where $g(T)$ is the temperature dependent coupling constant, running according 
to the two-loop  $\beta$-function, and $\gamma$ is a constant;  
$\gamma = 0.586 \pm 0.045$ for SU(3)~\cite{Karsch:1994af}.
In view of this, and for a better understanding of the nonperturbative structure
of QCD at high-$T$, a quantitative description of the properties of the spatial 
string tension is needed and the study of the Wilson connected correlator, 
lying in the spatial sublattice, at nonzero temperature provides us with an 
indirect measurement of $\sigma_s$.
\begin{figure}[tb] 
\centering
\includegraphics[width=0.48\textwidth,height=6.25cm]{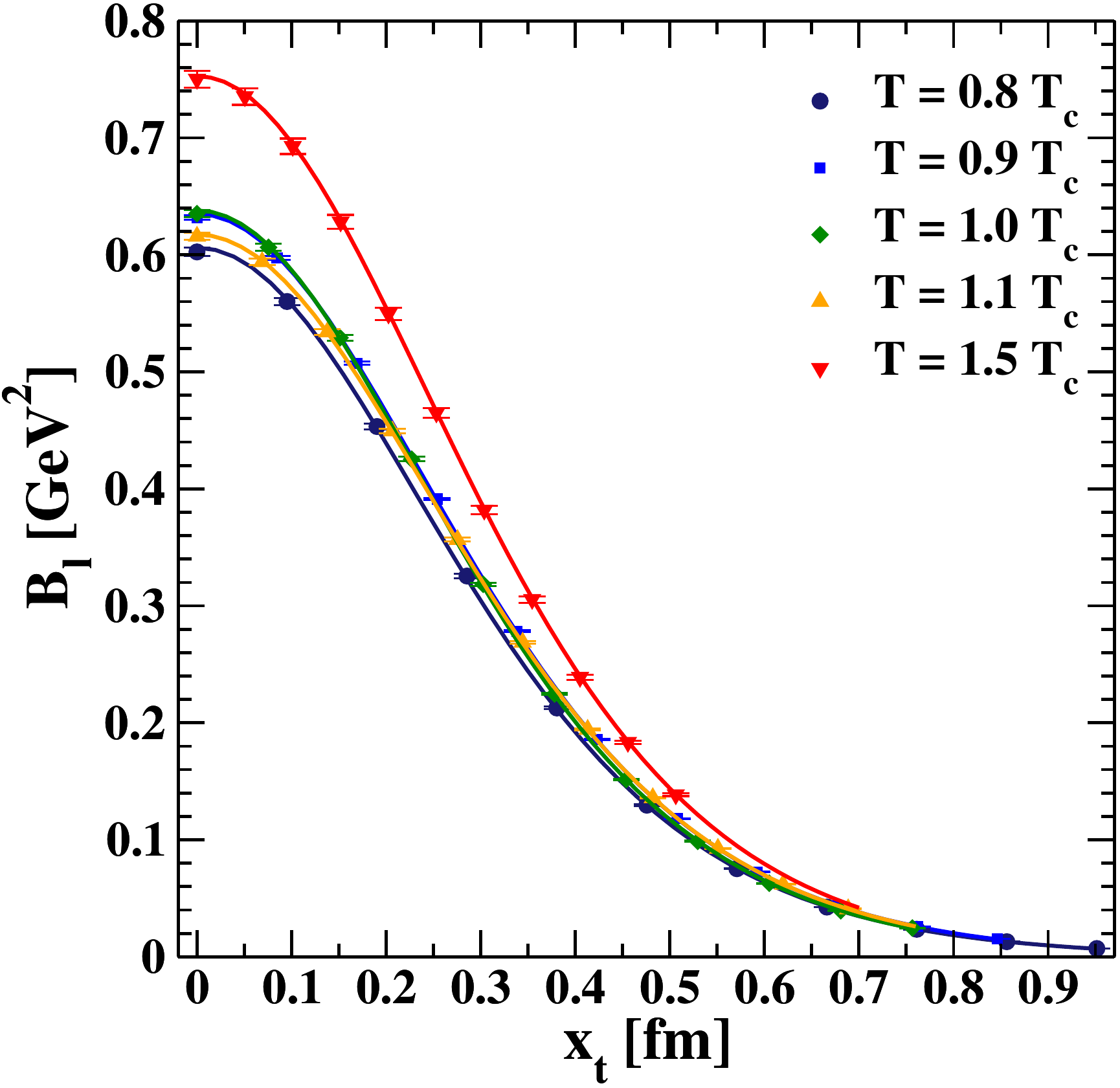}
\caption{$B_l\left(x_t\right)$ across the deconfinement transition, as 
determined from $\rho_W^{\rm conn}$ built in the spatial sublattice. The solid lines are the fit of our data to 
Eq.~\protect\eqref{eq:clem}.}
\label{fig:B_vs_xt_phys}
\end{figure}
Our results for the longitudinal chromomagnetic field {\it vs} $x_t$
are shown in Fig.~\ref{fig:B_vs_xt_phys} and the evidence is that our data are 
well fitted by Eq.~\eqref{eq:clem} at all temperatures and the field 
shape changes with $T$ in a way consistent with a growing spatial string.
\vspace{-0.35cm}
\section*{Acknowledgments}\vspace{-0.3cm}
This work was in part based on the MILC collaboration's public lattice gauge 
theory code. See~\url{http://physics.utah.edu/~detar/milc.html} and has 
been partially supported by the INFN - SUMA project. Simulations have been 
performed on BlueGene/Q at CINECA (CINECA - INFN agreement)\vspace{-0.1cm}

\vspace{-0.3cm}
\small{
}

\end{document}